\documentclass{article}
\usepackage{amssymb}
\usepackage{amsfonts}
\usepackage{amsmath}

\input{tcilatex}

\begin{document}

\title{An Irrotational Flow Field That Approximates Flat Plate Boundary
Conditions}
\author{Anthony A. Ruffa \\
Naval Undersea Warfare Center Division\\
1176 Howell Street\\
Newport, RI 02841}
\maketitle

\begin{abstract}
A irrotational solution is derived for the steady-state Navier-Stokes
equations that approximately satisfies the boundary conditions for flow over
a finite flat plate. \ The nature of the flow differs substantially from
boundary layer flow, with severe numerical difficulties in some regions.
\end{abstract}

\bigskip

\bigskip An analytic function having the form

\begin{equation}
f(z)=v+iu,
\end{equation}

leads to an exact solution of the two-dimensional steady-state Navier-Stokes
equations, i.e.,

\begin{eqnarray}
\rho _{0}\boldsymbol{u}\cdot \nabla \boldsymbol{u} &=&-\nabla p+\mu \nabla
^{2}\boldsymbol{u};  \notag \\
\nabla \cdot \boldsymbol{u} &=&0.
\end{eqnarray}

This occurs because (1) leads to a velocity field having the properties

\begin{equation}
\nabla ^{2}\boldsymbol{u=0};
\end{equation}

and

\begin{eqnarray}
\nabla \cdot \boldsymbol{u} &=&0;  \notag \\
\nabla \times \boldsymbol{u} &=&\boldsymbol{0}.
\end{eqnarray}

This is a subset of the generalized Beltrami flows$^{1}$, and note that (4)
are the Cauchy-Riemann equations for (1). \ Substituting (3) and (4) into
(2), and using the identity

\begin{equation}
\boldsymbol{u}\cdot \nabla \boldsymbol{u=}\left( \nabla \times \boldsymbol{u}%
\right) \times \boldsymbol{u}+\frac{1}{2}\nabla u^{2}
\end{equation}

leads to

\begin{equation}
\frac{\rho _{0}}{2}\nabla u^{2}=-\nabla p,
\end{equation}

or

\begin{equation}
\frac{p}{\rho _{0}}+\frac{u^{2}}{2}=C\text{.}
\end{equation}

The main difficulty with (1) involves finding a function $f(z)$ satisfying
useful no-slip boundary conditions. \ Consider the function

\begin{equation}
f(z)=\lim_{\epsilon \rightarrow 0}\frac{1}{L}\int_{0}^{L}\frac{\sin \left(
2\pi \left( z-z_{0}\right) /L\right) idx_{0}}{A+\left[ B\sin \left( 2\pi
\left( z-z_{0}\right) /L\right) -A\right] e^{\left( z-z_{0}\right)
^{2}/\epsilon }}.
\end{equation}

Here $z=x+iy$, $z_{0}=x_{0}+iy_{0}$ and $y_{0}=0$. \ When $y\rightarrow
\infty $,

\begin{equation}
f(z)\rightarrow \frac{1}{L}\int_{0}^{L}\frac{\sin \left( 2\pi \left(
z-z_{0}\right) /L\right) idx_{0}}{A}=0.
\end{equation}%
Making use of the identity

\begin{eqnarray}
\sin \left( 2\pi \left( z-z_{0}\right) /L\right) &=&\cos \left( 2\pi
x_{0}/L\right) \sin \left( 2\pi z/L\right) \\
&&-\sin \left( 2\pi x_{0}/L\right) \cos \left( 2\pi z/L\right) ,  \notag
\end{eqnarray}%
the terms $\cos \left( 2\pi z/L\right) $ and $\sin \left( 2\pi z/L\right) $
become large as $y\rightarrow \infty $; however, $\sin \left( 2\pi
x_{0}/L\right) $ and $\cos \left( 2\pi x_{0}/L\right) $ are precisely zero
when integrated over $0\leq x_{0}\leq L$.

When $y=0$,

\begin{equation}
f(z)=\lim_{\epsilon \rightarrow 0}\frac{1}{L}\int_{0}^{L}\frac{\sin \left(
2\pi \left( x-x_{0}\right) /L\right) idx_{0}}{A+\left[ B\sin \left( 2\pi
\left( x-x_{0}\right) /L\right) -A\right] e^{\left( x-x_{0}\right)
^{2}/\epsilon }}.
\end{equation}

When $x\neq x_{0}$, the denominator in (11) diverges. \ However, as $%
x\rightarrow x_{0}$, the denominator approaches $B\sin \left( 2\pi \left(
x-x_{0}\right) /L\right) $, so that as $\epsilon \rightarrow 0$, $u$
approaches a rectangle function between $x=0$ and $x=L$ and $v$ approaches
zero, meeting the boundary conditions for flow over a flat plate.

Although the integral (8) can be difficult to evaluate in some regions,
software packages seem to have less trouble converging when $B>>A$, at least
for $y=0$. \ Choosing $\epsilon =0.0001$, $A=5.6\times 10^{-4}$, $B=100A$
and $L=1\unit{m}$ satisfies the flat plate boundary condition as shown in
figure 1. \ As $\epsilon $ $\rightarrow 0$, $u$ becomes more step-like at $%
x=0$ and $x=1$. \ Numerical approaches require $\epsilon $ to be finite,
leading to regions at the plate edges where the boundary conditions are not
satisfied. \ These regions can be made as small as required by reducing $%
\epsilon $.

The boundary layer assumptions, i.e., $\frac{\partial ^{2}u}{\partial y^{2}}%
>>\frac{\partial ^{2}u}{\partial x^{2}}$ and $\frac{\partial p}{\partial y}%
=0 $ are useful for analyzing flow over a flat plate, but they break down in
the region at the leading edge$^{2}$, when $U_{0}x/\nu \lesssim 10000$. \
This region of non-validity can be used to define an acceptable region where
the flat plate boundary conditions are only satisfied approximately by (11)
and thus determines a practical value for $\epsilon $. \ When $U_{0}=1\unit{m%
}/\unit{s}$ and $\nu =10^{-6}\unit{m}^{2}/\unit{s}$, the extent of
non-validity is approximately defined by $x\lesssim 0.01\unit{m}$. \ Setting 
$\epsilon $ $=0.00001,$ $A=5.9\times 10^{-5}$, and $B=100A$ leads to the
boundary conditions being satisfied to within a tolerance on the order of $%
10^{-6}$ at the edges of the regions $-0.01\leq x\leq 0.01$ and $0.99\leq
x\leq 1.01$ as shown in figure 2.

\FRAME{ftbpFU}{4.4278in}{3.0364in}{0pt}{\Qcb{The $u$ approximation to the
flat plate boundary condition at $y=0$ for $\protect\epsilon =0.0001$.}}{}{%
Figure}{\special{language "Scientific Word";type
"GRAPHIC";maintain-aspect-ratio TRUE;display "USEDEF";valid_file "T";width
4.4278in;height 3.0364in;depth 0pt;original-width 9.487in;original-height
6.487in;cropleft "0";croptop "1";cropright "1";cropbottom "0";tempfilename
'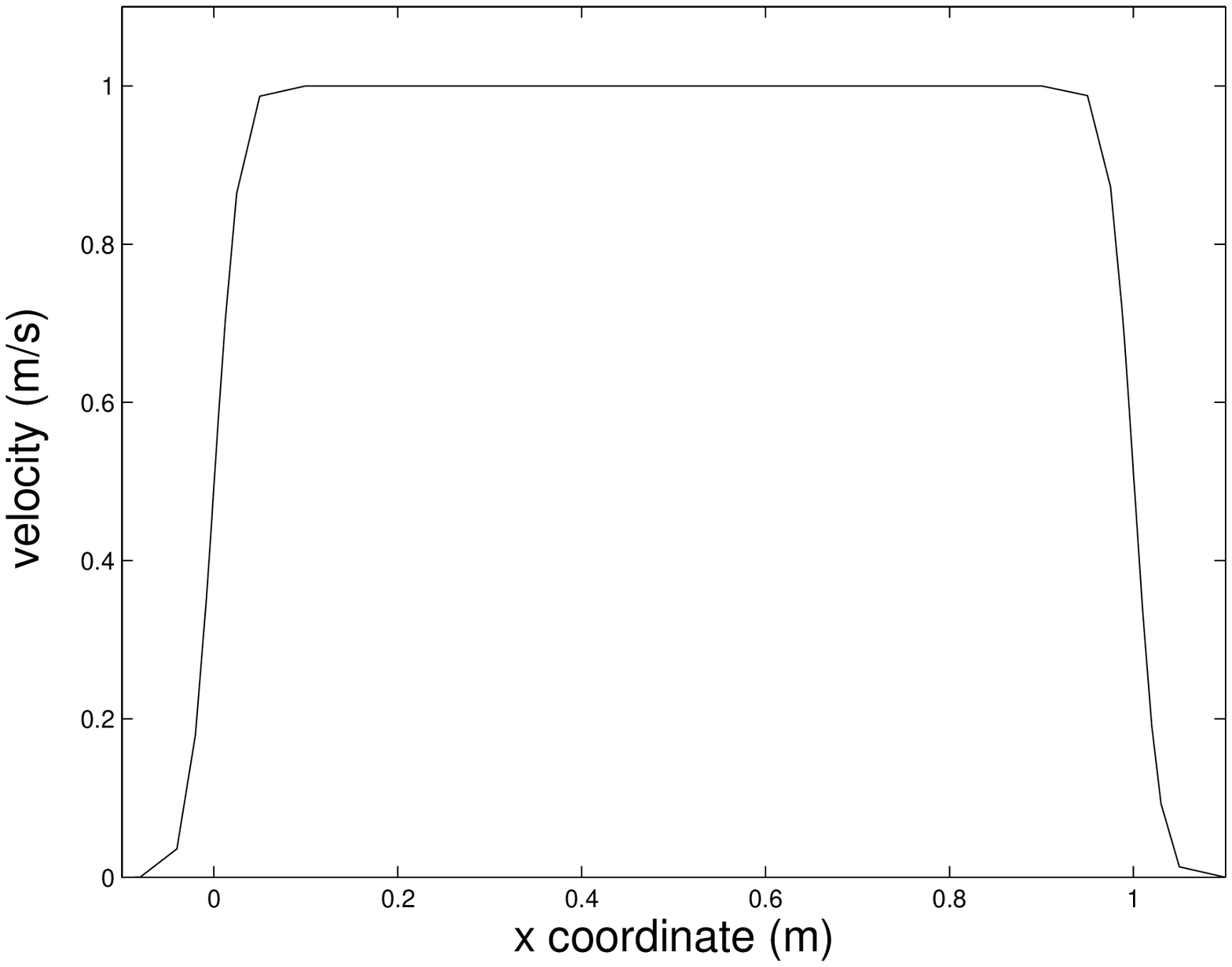';tempfile-properties "XPR";}}

\FRAME{ftbpFU}{4.4278in}{3.0364in}{0pt}{\Qcb{The $u$ approximation to the
flat plate boundary condition at $y=0$ for $\protect\epsilon =0.00001$.}}{}{%
Figure}{\special{language "Scientific Word";type
"GRAPHIC";maintain-aspect-ratio TRUE;display "USEDEF";valid_file "T";width
4.4278in;height 3.0364in;depth 0pt;original-width 9.487in;original-height
6.487in;cropleft "0";croptop "1";cropright "1";cropbottom "0";tempfilename
'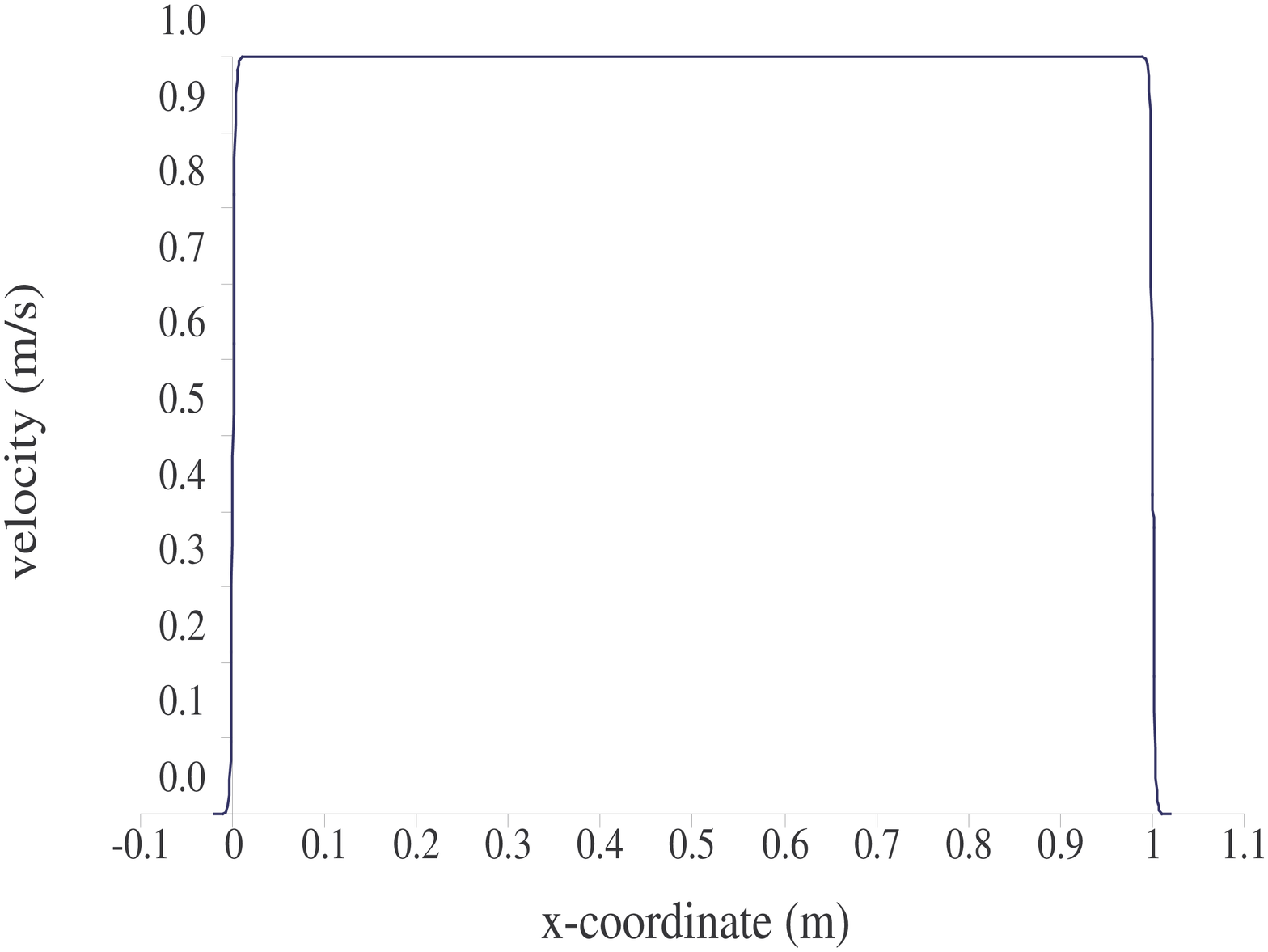';tempfile-properties "XPR";}}

Although the flow described by (8) exactly satisfies the boundary conditions
for a finite flat plate when $\epsilon \rightarrow 0$ (and approximately for
small but finite $\epsilon $), the flow field differs substantially from
boundary layer flow. \ Figure 3 shows the regions having nonzero velocity
components. \ As $y$ increases, the integrand evolves into a sinusoidal-like
function, until it becomes sufficiently sinusoidal that $u$ and $v$ both
approach zero. \ Just below the upper "zero" line, small perturbations in
the sinusoidal-like integrand functions lead to nonzero velocities. \ In
some regions, either the first term or the second in the denominator in (8)
dominate, allowing the other to be neglected. \ The latter is true when $%
y<<1 $ above the plate, allowing the velocity field to be accurately
approximated. \ When neither terms can be neglected, evaluation of the
integral becomes tedious because of very large integrand magnitudes, leading
software packages to either fail to converge or converge to incorrect
solutions. \ Numerical determination of the velocity field for many regions
remains a challenge.

\FRAME{ftbpFU}{5.4068in}{4.0629in}{0pt}{\Qcb{Upper and lower "zero" lines.
All nonzero velocity components are confined to the region between these
lines.}}{}{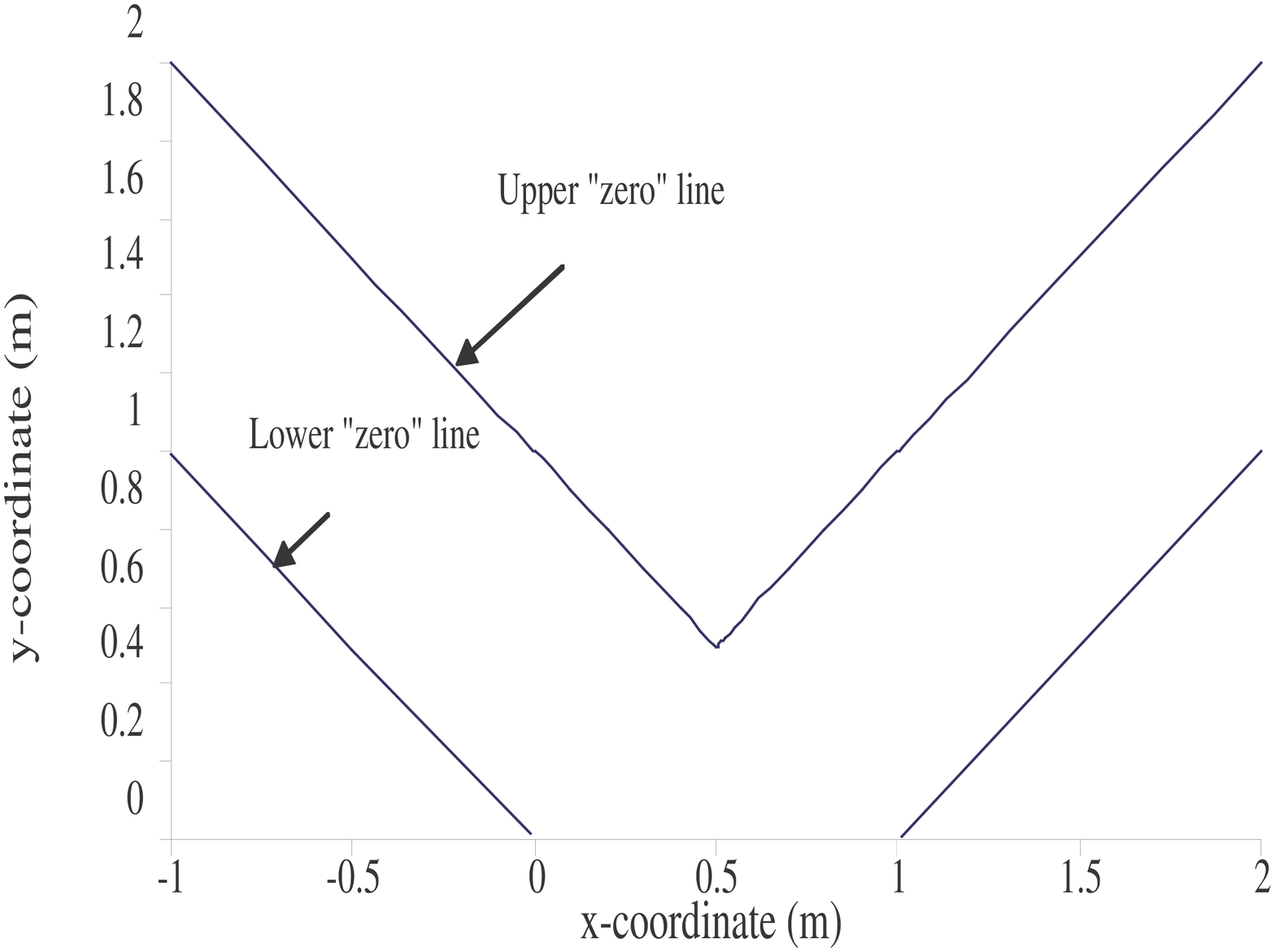}{\special{language "Scientific Word";type
"GRAPHIC";maintain-aspect-ratio TRUE;display "USEDEF";valid_file "F";width
5.4068in;height 4.0629in;depth 0pt;original-width 9.9998in;original-height
7.4996in;cropleft "0";croptop "1";cropright "1";cropbottom "0";filename
'tony3.ps';file-properties "XNPEUR";}}

\bigskip

\bigskip

\bigskip

\bigskip

\bigskip

\bigskip

\bigskip

\bigskip

\bigskip

\bigskip

\bigskip

\bigskip

\bigskip

\bigskip

\bigskip

\bigskip

\bigskip

\bigskip

\bigskip

\bigskip

\bigskip

\bigskip

\bigskip

\bigskip

\bigskip

\bigskip

\bigskip

\bigskip

\bigskip

\bigskip

\bigskip

\bigskip

\bigskip

\bigskip

\bigskip

\bigskip

\textbf{References}

1. \ Wang, C. Y. 1991. "Exact Solutions of the Steady-State Navier-Stokes
Equations." \ \textit{Annu. Rev. Fluid Mech.} \textbf{23}, 159-177.

2. \ Schlichting, H., "Boundary-Layer Theory", 6th ed., McGraw-Hill, 1968.

\bigskip

\bigskip

\bigskip

\bigskip

\bigskip

\bigskip

\bigskip

\bigskip

\bigskip

\bigskip

\bigskip

\bigskip

\bigskip

\bigskip

\bigskip

\end{document}